\documentclass[%
reprint,
showpacs,preprintnumbers,
amsmath,amssymb,
prl
]{revtex4-1}

\usepackage{tikz}
\usepackage{pgffor}
\usepackage{verbatim}
\usepackage{pstricks}
\usepackage{pst-3dplot}
\usepackage[colorlinks, breaklinks=true, linkcolor=purple, citecolor=purple, linktocpage=true]{hyperref}
\usepackage{color}
\usepackage{graphicx}
\usepackage{dcolumn}
\usepackage{bm}

\usepackage{txfonts}

\begin{document}

\title{Chiral Magnetic Effect and Anomalous Hall Effect in Antiferromagnetic Insulators\\ with Spin-Orbit Coupling}

\author{Akihiko Sekine}
\email{sekine@imr.tohoku.ac.jp}
\author{Kentaro Nomura}
\affiliation{Institute for Materials Research, Tohoku University, Sendai 980-8577, Japan}

\date{\today}

\begin{abstract}
We search for dynamical magnetoelectric phenomena in three-dimensional correlated systems with spin-orbit coupling.
We focus on the antiferromagnetic insulator phases where the dynamical axion field is realized by the fluctuation of the antiferromagnetic order parameter.
It is shown that the dynamical chiral magnetic effect, an alternating current generation by magnetic fields, emerges due to the time dependences of the order parameter such as antiferromagnetic resonance.
It is also shown that the anomalous Hall effect arises due to the spatial variations of the order parameter such as antiferromagnetic domain walls.
Our study indicates that spin excitations in antiferromagnetic insulators with spin-orbit coupling can result in nontrivial charge responses.
Moreover, observing the chiral magnetic effect and anomalous Hall effect in our system is equivalent to detecting the dynamical axion field in condensed matter.
\end{abstract}


\maketitle

{\it Introduction.---}
Antiferromagnets (AFMs) have attracted much attention from the viewpoints of both purely scientific and applied research.
Theoretically, it is known that antiferromagnetic (AF) phases are often favored in systems with strong on-site repulsive interactions.
In the vicinity of AF phases, the emergence of exotic phases and phenomena such as high-temperature superconductivity and spin liquids is well acknowledged \cite{Imada1998,Dagotto2005,Balents2010}.
On the other hand, AFMs have recently been studied intensively in the field of spintronics \cite{Cheng2014,Wang2014,Hahn2014,Moriyama2015}, as a possible new class of materials alternative to ferromagnets.
These studies suggest that the staggered magnetization can play essential roles, although AFMs had not been considered suitable for practical use due to the lack of net magnetization unlike ferromagnets.

Recent intensive and extensive studies have revealed the importance of spin-orbit coupling (SOC) in condensed matter.
Especially, the discovery that strong SOC is essential to realize topologically nontrivial phases opened a new direction in modern physics \cite{Hasan2010,Qi2011,Ando2013}.
Since topological invariants are determined from electronic band structures, studies of topological phases started as a single-particle problem.
Subsequently many-body effects in topological phases and spin-orbit coupled systems have become an attractive subject \cite{Hohenadler2013,Witczak-Krempa2014}, and the emergence of novel phases such as the topological Mott insulator \cite{Pesin2010} (or more generally the fractionalized topological insulators \cite{Maciejko2015}) and the Weyl semimetal \cite{Wan2011} has been predicted.
As for novel phenomena in spin-orbit coupled and correlated systems, for example, the axionic polariton, a total reflection phenomenon of light, has been suggested \cite{Li2010}.

In this Letter, we study electromagnetic responses of antiferromagnetic insulator (AFI) phases in three-dimensional (3D) correlated systems with SOC.
We explore dynamical magnetoelectric phenomena where the staggered magnetization plays essential roles.
We show that, in the presence of SOC, spin excitations in AFIs can result in nontrivial {\it charge} responses, as a consequence of the realization of the dynamical axion field.
First we show the emergence of the chiral magnetic effect (CME), an electric current generation by magnetic fields \cite{Fukushima2008}, in the AFI phase.
The CME was originally proposed in gapless Dirac fermion systems \cite{Fukushima2008}, and its possibility has been discussed in Weyl semimetals \cite{Zyuzin2012,Vazifeh2013,Goswami2013,Burkov2015,Chang2015,Buividovich2015}.
In contrast to preceding works, we propose the {\it dynamical} realization of the CME in {\it gapped} systems.
We also show the occurrence of the anomalous Hall effect (AHE) in the AFI phase.
It is known that the AHE occurs usually in ferromagnetic metals \cite{Nagaosa2010}, while the AHE arising from nontrivial spin textures has been studied in frustrated or noncollinear AFMs \cite{Shindou2001,Chen2014}.
We propose that spatial variations of the staggered magnetization lead to the AHE.

{\it Realization of the Dynamical Axion Field and its Consequences.---}
Let us consider 3D electron systems having both on-site interactions and SOC, such as $5d$ transition metal oxides \cite{Witczak-Krempa2014,Pesin2010,Jackeli2009}.
We focus on systems that become magnetically ordered Mott insulators when on-site interactions are strong, while they are topological band insulators when on-site interactions are weak.
Once a magnetic order is formed, the mean-field approximation of the interaction term can capture the essential physics of the system.
In this work, we particularly consider AFIs whose mean-field lattice Hamiltonian is given by
$\mathcal{H}(\bm{k})=\epsilon_0(\bm{k})\bm{1}+\sum_{\mu=1}^5R_\mu(\bm{k})\alpha_\mu$.
Here $\bm{k}=(k_1,k_2,k_3)$ is a wave vector in the Brillouin zone, $\bm{1}$ is the $4\times 4$ identity matrix, and  the $4\times 4$ matrices $\alpha_\mu$ satisfy the Clifford algebra $\{\alpha_\mu,\alpha_\nu\}=2\delta_{\mu\nu}$ with $\alpha_5=\alpha_1\alpha_2\alpha_3\alpha_4$.
The Hamiltonian of this form can be realized, for example, in the AFI phases of Bi$_2$Se$_3$ family doped with magnetic impurities such as Fe \cite{Li2010} and transition metal oxides with corundum structure such as $\alpha$-Fe$_2$O$_3$ \cite{Wang2011}.
In this case, we can derive 3D massive Dirac Hamiltonians of the form
\begin{align}
\begin{split}
\mathcal{H}_{\rm eff}(\bm{q})=q_1\alpha_1+q_2\alpha_2+q_3\alpha_3+M_0\alpha_4+M_{5f}\alpha_5\label{alpha5-Dirac}
\end{split}
\end{align}
around some momentum points $X_f$, where $\bm{q}=\bm{k}-X_f$.
The subscript $f$ indicates the valley degrees of freedom.
The kinetic term $\sum_{\mu=1}^3q_\mu\alpha_\mu$ is spin-dependent as a consequence of SOC.
$M_0\alpha_4$ is a mass term with time-reversal and parity (spatial inversion) symmetries induced by SOC, and $M_{5f}\alpha_5$ is a mass term with broken time-reversal and parity symmetries induced by mean-field AF order parameter.
We require that the system is a topological insulator when $M_0>0$ and $M_{5f}=0$.

In what follows, we consider consequences arising from the existence of the $M_{5f}\alpha_5$ mass term.
The effective action of the system in the presence of an external electromagnetic potential $A_\mu$ is written as
\begin{align}
\begin{split}
S_{\rm eff}=\int dtd^3 r\sum_f\bar{\psi}_f(\bm{r},t)\left[i\gamma^\mu D_\mu-M'_f e^{i\theta_f\gamma^5}\right]\psi_f(\bm{r},t),\label{Eff-Action}
\end{split}
\end{align}
where $t$ is real time, $\psi_f(\bm{r},t)$ is a four-component spinor, $\bar{\psi}_f=\psi_f^\dagger\gamma^0$, $D_\mu=\partial_\mu+ieA_\mu$, $M'_f=\sqrt{(M_0)^2+(M_{5f})^2}$, $\cos\theta_f=M_0/M'_f$, $\sin\theta_f=-M_{5f}/M'_f$, and we have used the fact that $\alpha_4=\gamma^0$, $\alpha_5=-i\gamma^0\gamma^5$ and $\alpha_j=\gamma^0\gamma^j$ ($j=1,2,3$).
By applying the Fujikawa's method \cite{Fujikawa1979} to the action (\ref{Eff-Action}), the theta term is obtained as \cite{comment-A}
\begin{align}
\begin{split}
S_\theta&=\int dtd^3 r \frac{e^2}{2\pi h}\theta \bm{E}\cdot\bm{B},
\label{S_theta_realtime}
\end{split}
\end{align}
where $\theta=\frac{\pi}{2}[1+\mathrm{sgn}(M_0)]-\sum_f\tan^{-1}(M_{5f}/M_0)$, and $\bm{E}$ ($\bm{B}$) is an external electric (magnetic) field.
From this action, we obtain the magnetoelectric responses expressed by $\bm{P}=\theta e^2/(2\pi h)\bm{B}$ and $\bm{M}=\theta e^2/(2\pi h)\bm{E}$, with $\bm{P}$ the electric polarization and $\bm{M}$ the magnetization.
In 3D time-reversal invariant topological (normal) insulators, $\theta=\pi$ ($\theta=0$) \cite{Qi2008}.
However, the value of $\theta$ can be arbitrary when time-reversal and parity symmetries of the system are broken \cite{Essin2009,Essin2010,Coh2011}.
Furthermore, when the value of $\theta$ depends on space and time, it can be said that the dynamical axion field is realized in condensed matter \cite{Li2010}.
Some consequences of the realization have been studied so far \cite{Li2010,Ooguri2012}.

Notice that, when the dynamical axion field is realized, the theta term can be rewritten in the Chern-Simons form as
\begin{align}
\begin{split}
S_\theta=-\int dt d^3r\frac{e^2}{4\pi h}\epsilon^{\mu\nu\rho\lambda}[\partial_\mu \theta(\bm{r},t)] A_\nu\partial_\rho A_\lambda.\label{Theta-term-Chern-Simons}
\end{split}
\end{align}
Then the induced four-current density $j^\nu$ can be obtained from the variation of the above action with respect to the four-potential $A_\nu$:
$
j^\nu=\frac{\delta S_\theta}{\delta A_\nu}=-\frac{e^2}{2\pi h}[\partial_\mu \theta(\bm{r},t)]\epsilon^{\mu \nu\rho\lambda}\partial_\rho A_\lambda.
$
The induced current density is given by \cite{Wilczek1987,comment-B}
\begin{align}
\begin{split}
\bm{j}(\bm{r},t)&=\frac{e^2}{2\pi h}\left[\dot{\theta}(\bm{r},t)\bm{B}+\nabla\theta(\bm{r},t)\times\bm{E}\right],\label{Current-Eq}
\end{split}
\end{align}
where $\dot{\theta}=\partial \theta(\bm{r},t)/\partial t$.
The magnetic-field-induced term is the CME \cite{Fukushima2008}.
The electric-field-induced term is the AHE, since it is perpendicular to the electric field.
The induced current of the form (\ref{Current-Eq}) has been also studied in Weyl semimetals \cite{Zyuzin2012,Vazifeh2013,Goswami2013,Burkov2015}, where the chemical potential difference between the band touching points and the separation of the points in momentum space are required for the CME and AHE, respectively.
However, the existence of the CME in Weyl semimetals is still being discussed theoretically \cite{Zyuzin2012,Vazifeh2013,Goswami2013,Burkov2015,Chang2015,Buividovich2015}.
Note that the situation we consider in this paper is completely different, since the system is gapped, i.e., the above conditions required in the case of Weyl semimetals are not needed.

{\it Theoretical Model.---}
To study the induced current (\ref{Current-Eq}) more concretely, let us consider a 3D lattice model with SOC and electron correlations.
The model we adopt is the Fu-Kane-Mele-Hubbard model on a diamond lattice at half-filling, whose Hamiltonian is given by \cite{Fu2007a,Fu2007,Sekine2014}
\begin{align}
\begin{split}
H&=\sum_{\langle i,j\rangle,\sigma}t_{ij}c^\dag_{i\sigma}c_{j\sigma}+i\frac{4\lambda}{a^2}\sum_{\langle\langle i,j\rangle\rangle}c^\dag_{i}\bm{\sigma}\cdot(\bm{d}^1_{ij}\times\bm{d}^2_{ij})c_{j}\\
&\quad+U\sum_i n_{i\uparrow}n_{i\downarrow},\label{FKMH-model}
\end{split}
\end{align}
where $c^\dag_{i\sigma}$ is an electron creation operator at a site $i$ with spin $\sigma(=\uparrow,\downarrow)$, $n_{i\sigma}=c^\dag_{i\sigma}c_{i\sigma}$, and $a$ is the lattice constant of the fcc lattice.
The first through third terms represent the nearest-neighbor hopping, the next-nearest-neighbor SOC, and the on-site electron-electron interaction, respectively.
$\bm{d}^1_{ij}$ and $\bm{d}^2_{ij}$ are the two vectors which connect two sites $i$ and $j$ on the same sublattice.
Namely they are given by two of the four nearest-neighbor bond vectors.
$\bm{\sigma}=(\sigma_1,\sigma_2,\sigma_3)$ are the Pauli matrices for the spin degrees of freedom.
We introduce a lattice distortion such that $t_{ij}=t+\delta t_1$ for the [111] direction and $t_{ij}=t$ for the other three directions, which induces a bandgap of $2|M_0|$ ($M_0\equiv \delta t_1$) in the noninteracting spectrum.

We perform the mean-field approximation to the interaction term as
$H_U=U\sum_i n_{i\uparrow}n_{i\downarrow}\approx  U\sum_i \left[ \langle n_{i\downarrow}\rangle n_{i\uparrow}+\langle n_{i\uparrow}\rangle n_{i\downarrow}-\langle n_{i\uparrow}\rangle\langle n_{i\downarrow}\rangle
-\langle c^\dagger_{i\uparrow}c_{i\downarrow}\rangle c^\dagger_{i\downarrow}c_{i\uparrow}-\langle c^\dagger_{i\downarrow}c_{i\uparrow}\rangle\right.$
$\left.\times c^\dagger_{i\uparrow}c_{i\downarrow}+\langle c^\dagger_{i\uparrow}c_{i\downarrow}\rangle \langle c^\dagger_{i\downarrow}c_{i\uparrow}\rangle \right]$.
SOC breaks spin SU(2) symmetry and the orientations of the spins are coupled to the lattice structure.
Hence we should parametrize the AF ordering between the two sublattices $A$ and $B$ in terms of the spherical coordinate $(n,\theta,\varphi)$:
\begin{align}
\begin{split}
\langle \bm{S}_{i'A}\rangle=-\langle \bm{S}_{i'B}\rangle&=(n\sin\theta\cos\varphi, n\sin\theta\sin\varphi, n\cos\theta)\\
&\equiv n_1\bm{e}_x+n_2\bm{e}_y+n_3\bm{e}_z\ (\equiv \bm{n}),\label{AF-order}
\end{split}
\end{align}
where $\langle \bm{S}_{i'\mu}\rangle=\frac{1}{2}\langle c^\dagger_{i'\mu\alpha}\bm{\sigma}_{\alpha\beta}c_{i'\mu\beta}\rangle$ $(\mu=A,B)$ with $i'$ denoting the $i'$-th unit cell.
In the following we consider the ground state given by $(n_0, \theta_0, \varphi_0)$.
The low-energy effective Hamiltonian of the AFI phase is written in the form (\ref{alpha5-Dirac}):
$H_{\rm eff}=\sum_{\bm{q}}\sum_{f=1,2,3}\psi^\dag_{f\bm{q}}\mathcal{H}_f(\bm{q})\psi_{f\bm{q}}$,
where $\psi_{f\bm{q}}$ is a four-component spinor \cite{Sekine2014}.
Therefore the value of $\theta$ in the AFI phase of the Fu-Kane-Mele-Hubbard model is given by \cite{Sekine2014}
\begin{align}
\begin{split}
\theta=\frac{\pi}{2}[1+\mathrm{sgn}(M_0)]-\sum_{f=1,2,3}\tan^{-1}\left(Un_f/M_0\right).\label{Chap6-Theta}
\end{split}
\end{align}
From this equation, we see that the dynamical axion field is realized by the fluctuation of the AF order parameter $n_f$, i.e., by the spin excitations \cite{Li2010,Wang2011}.

{\it Dynamical Chiral Magnetic Effect.---}
\begin{figure}[!t]
\centering
\includegraphics[width=0.85\columnwidth,clip]{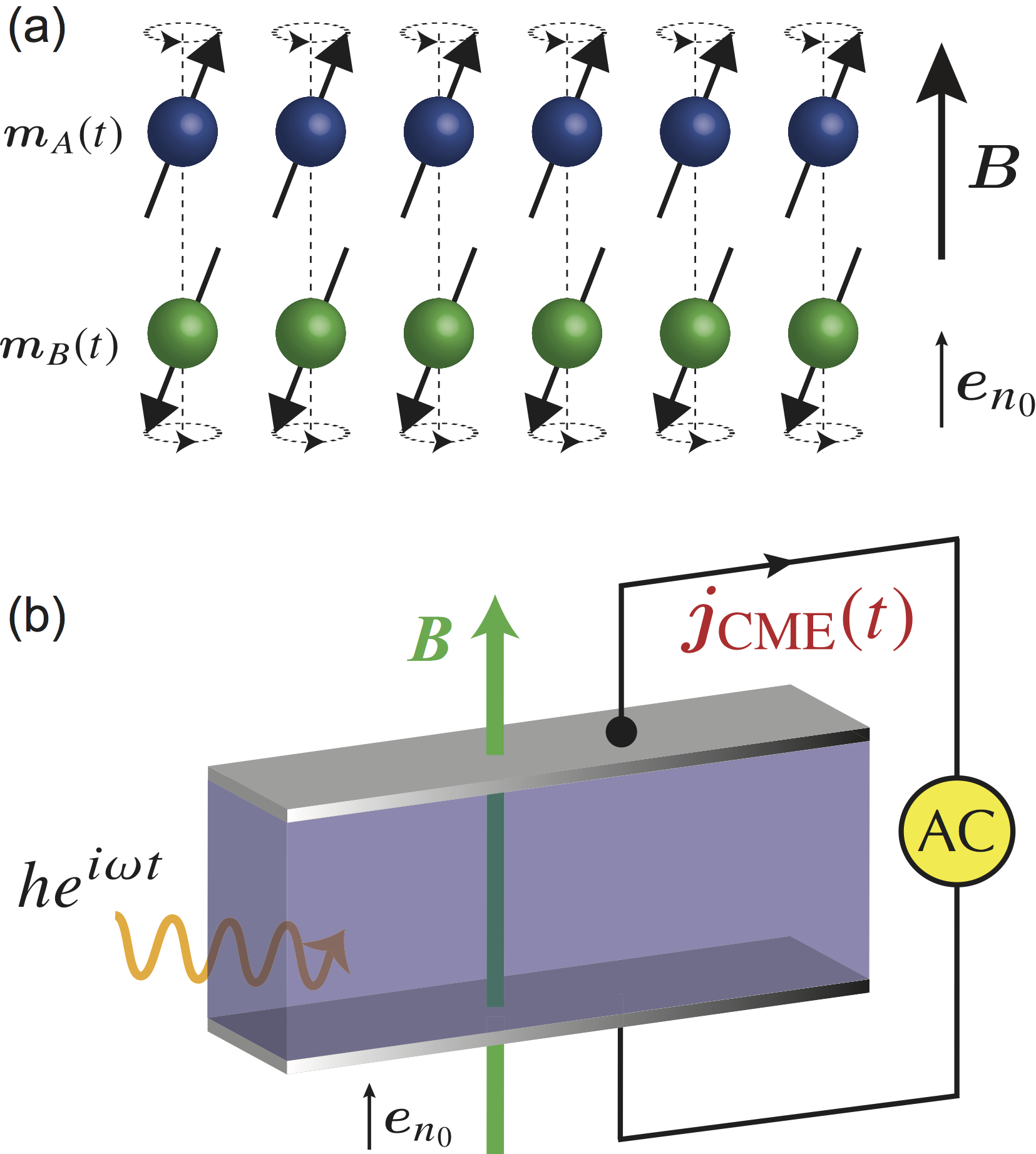}
\caption{(Color online) (a) Schematic figure of the dynamics of $\bm{m}_A$ and $\bm{m}_B$ in the AF resonance state.
(b) A possible experimental setup to observe the CME in our system.
A static magnetic field $\bm{B}$ is applied along the easy axis of the AF order, $\bm{e}_{n_0}=\bm{n}_0/|\bm{n}_0|$.
When the external microwave (i.e., ac magnetic field) frequency $\omega$ is equal to the resonance frequencies $\omega_\pm$, the induced alternating current $\bm{j}_{\rm CME}(t)$ will be observed.}\label{Fig1}
\end{figure}
First we focus on the magnetic-field-induced term in Eq. (\ref{Current-Eq}), i.e., the CME in the AFI phase:
\begin{align}
\begin{split}
\bm{j}_{\rm CME}(\bm{r},t)=-\frac{e^2}{2\pi h}\sum_f\frac{UM_0}{M_0^2+(Un_f)^2}\dot{n}_f(\bm{r},t)\bm{B}.\label{DCME}
\end{split}
\end{align}
Let us consider a case where a microwave (i.e., ac magnetic field) is irradiated and a static magnetic field $\bm{B}=B\bm{e}_{n_0}$ is applied along the easy axis of the AF order.
Here $\bm{e}_{n_0}=\bm{n}_0/|\bm{n}_0|$ is the unit vector parallel to the easy axis.
The dynamics of the sublattice magnetizations $\langle \bm{S}_{i'A}\rangle=\bm{m}_A$ and $\langle \bm{S}_{i'B}\rangle=\bm{m}_B$ can be described by \cite{Keffer1952,comment1}
\begin{align} 
\begin{split}
\dot{\bm{m}}_A&=\bm{m}_A\times\left\{-\omega_J\bm{m}_B+\left[g\mu_B B+\omega_A(\bm{m}_A\cdot\bm{e}_{n_0})\right]\bm{e}_{n_0}\right\},\\
\dot{\bm{m}}_B&=\bm{m}_B\times\left\{-\omega_J\bm{m}_A+\left[g\mu_B B+\omega_A(\bm{m}_B\cdot\bm{e}_{n_0})\right]\bm{e}_{n_0}\right\},
\end{split}\label{dynamics}
\end{align}
where $\omega_J$ and $\omega_A$ are the exchange field and anisotropy field, respectively.
We write $\bm{m}_A$ and $\bm{m}_B$ as $\bm{m}_A=n_0\bm{e}_{n_0}+\delta\bm{m}_{A\bot} e^{i\omega t}$ and $\bm{m}_B=-n_0\bm{e}_{n_0}+\delta\bm{m}_{B\bot} e^{i\omega t}$.
Up to linear order in $\delta\bm{m}_{A(B)\bot}$ (i.e., $|\delta\bm{m}_{A(B)\bot}|\ll 1$), we obtain the resonance frequencies as $\omega=\omega_\pm=g\mu_B B\pm\sqrt{(2\omega_J+\omega_A)\omega_A}$.
In the resonance state, where all the spins are precessing around the easy axis with the same frequency $\omega_+$ (or $\omega_-$), the AF order parameter is described as
\begin{align}
\begin{split}
\bm{n}_\pm(t)\equiv[\bm{m}_A(t)-\bm{m}_B(t)]/2\approx n_0\bm{e}_{n_0}+\delta\bm{n}_\pm e^{i\omega_\pm t}\label{m(t)}.
\end{split}
\end{align}
Here we have neglected the difference between the angles $\theta_A=\tan^{-1}(|\delta\bm{m}_{A\bot}|/n_0)$ and $\theta_B=\tan^{-1}(|\delta\bm{m}_{B\bot}|/n_0)$.
The ratio $\theta_A/\theta_B$ is obtained as $\theta_A/\theta_B\approx(1+\sqrt{\omega_A/\omega_J})^2$ \cite{Keffer1952}.
Typically, the ratio $\omega_A/\omega_J(\approx K/J)$, with $K$ and $J$ being the strength of the anisotropy and exchange coupling, respectively, is of the order of $10^{-2}$ to $10^{-3}$ \cite{Foner1963}.
Therefore we see that $\bm{m}_A\approx -\bm{m}_B$ and thus $\dot{\bm{m}}_A\approx -\dot{\bm{m}}_B$.
Illustration of the dynamics of $\bm{m}_A$ and $\bm{m}_B$ in the AF resonance state is shown in Fig. \ref{Fig1}(a).

From the relation such that $\bm{n}=n_1\bm{e}_x+n_2\bm{e}_y+n_3\bm{e}_z$, 
we have $n_1=n_0\sin\theta_0\cos\varphi_0+\delta n\cos\omega t\cos\theta_0\cos\varphi_0-\delta n\sin\omega t\sin\varphi_0$,
$n_2=n_0\sin\theta_0\sin\varphi_0+\delta n\cos\omega t\cos\theta_0\sin\varphi_0+\delta n\sin\omega t\cos\varphi_0$, and
$n_3=n_0\cos\theta_0-\delta n\cos\omega t\sin\theta_0$.
Substituting these quantities into Eq. (\ref{DCME}), we obtain the analytical expression for $\bm{j}_{\rm CME}(t)$.
Especially, in the vicinity of the phase boundary where $Un_f/M_0\ll 1$ \cite{comment2}, Eq. (\ref{DCME}) is simplified as
\begin{align}
\begin{split}
\bm{j}_{\rm CME}(t)
=\frac{e^2}{2\pi h}\frac{UD_1}{M_0}\bm{B}\sum_{a=\pm}\omega_a\delta n_a\sin\left(\omega_a t+\alpha\right),\label{expression-j-CME}
\end{split}
\end{align}
where $D_1=\sqrt{p^2+q^2}$ and $\tan\alpha=q/p$ with $p=(\cos\varphi_0+\sin\varphi_0)\cos\theta_0-\sin\theta_0$ and $q=\sin\varphi_0-\cos\varphi_0$.
Equation (\ref{expression-j-CME}) means that an alternating current is induced by the AF resonance.
Schematic figure of a possible experimental setup to observe the CME in our system is shown in Fig. \ref{Fig1}(b).
$\delta n_\pm$ is a function of the external microwave frequency $\omega$ with Lorentzian structure, i.e., $\delta  n_\pm(\omega)\sim a/[(\omega-\omega_\pm)^2+a^2]$ with $a$ being a constant.
Therefore two peaks will appear in the intensity $|\bm{j}_{\rm CME}(\omega)|$.

Here let us estimate the maximum value of the CME (\ref{expression-j-CME}): $|j_{\rm CME}|_{\rm max}=\frac{e^2}{2\pi h}\frac{U|D_1|}{|M_0|}B\omega_\pm\delta n_\pm$.
Substituting possible values $Un_0/|M_0|\sim 0.1$ \cite{comment3}, $|D_1|\sim 1$, $\delta n_\pm/n_0\sim 0.02$, and $\omega_\pm\sim 500\ \mathrm{GHz}$ at $B\sim 1\ \mathrm{T}$ \cite{Nagata1974,Hagiwara1999}, we have $|j_{\rm CME}|_{\rm max}\sim 1\times 10^4\ \mathrm{A/m^2}$.
This value is experimentally observable.
It should be noted that the current is adiabatically induced in the gapped phase as in the case of the quantum Hall effect or the topological charge pumping effect \cite{Thouless1983}.
Hence there is no energy dissipation, unlike the conventional transport regime which causes the Joule heat.

The CME was originally proposed in the ground states of massless Dirac fermion systems as a direct current generation by static magnetic fields due to the presence of the chemical potential difference between band touching points \cite{Fukushima2008}.
If such a static CME exists in realistic materials, there will be substantial potentials for its applications, since the current is dissipationless.
However, the existence of the CME remains a theoretically controversial subject in Weyl semimetals \cite{Zyuzin2012,Vazifeh2013,Goswami2013,Burkov2015,Chang2015,Buividovich2015}.
As discussed in Ref. \onlinecite{Vazifeh2013}, the possibility of the static CME would be ruled out in crystalline solids (i.e., lattice systems), regardless of the presence or absence of energy gaps.
In contrast to preceding works, our study proposes that the CME occurs {\it dynamically} in insulating systems, which requires time dependences of the AF order parameter $n_f$ caused by external forces such as AF resonance state.

{\it Anomalous Hall Effect.---}
\begin{figure}[!t]
\centering
\includegraphics[width=1.0\columnwidth,clip]{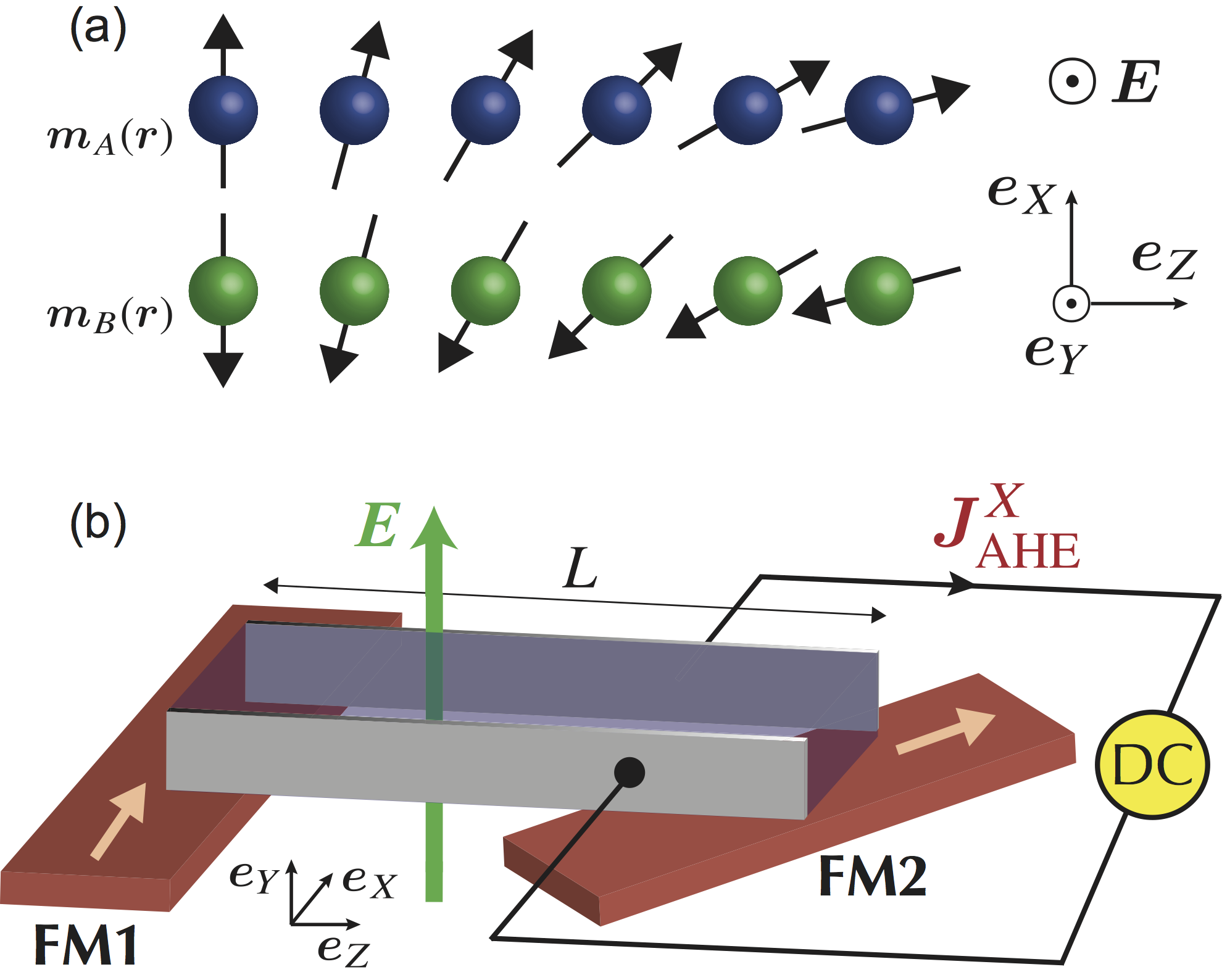}
\caption{(Color online) (a) Schematic figure of a one-dimensional AF texture, an orientational domain wall of length $L$.
The AF order parameter $\bm{n}(\bm{r})=[\bm{m}_A(\bm{r})-\bm{m}_B(\bm{r})]/2$ at the two edges has a relative angle $\delta$.
(b) A possible experimental setup to observe the AHE in our system.
A static electric filed $\bm{E}$ is applied perpendicular to the AF order.
There is a relative angle $\delta$ in the magnetization directions of ferromagnet 1 (FM1) and ferromagnet 2 (FM2).}\label{Fig2}
\end{figure}
Next we focus on the electric-field-induced term in Eq. (\ref{Current-Eq}), i.e., the AHE in the AFI phase:
\begin{align}
\begin{split}
\bm{j}_{\rm AHE}(\bm{r},t)=-\frac{e^2}{2\pi h}\sum_f\frac{UM_0}{M_0^2+(Un_f)^2}\nabla n_f(\bm{r},t)\times\bm{E}.\label{AHE}
\end{split}
\end{align}
In order to obtain a concrete expression for this current, let us move on to a new Cartesian coordinate $(X, Y, Z)$.
We consider a one-dimensional AF texture of length $L$ along the $Z$ direction, an orientational domain wall \cite{Bode2006,Tveten2013}.
As shown in Fig. \ref{Fig2}(a), the AF order parameter $\bm{n}(\bm{r})=[\bm{m}_A(\bm{r})-\bm{m}_B(\bm{r})]/2$ at the two edges has a relative angle $\delta$.
Namely, we have $\theta(Z=0)=\theta_0$ and $\theta(Z=L)=\theta_0+\delta$ in the original spherical coordinate.
A static electric filed $\bm{E}$ is applied perpendicular to the AF order as $\bm{E}=E_Y\bm{e}_{Y}$.
For simplicity, we assume that the system lies near the phase boundary where $Un_f/M_0\ll 1$ \cite{comment2}.
Noting that only the $X$ component $j^X_{\rm AHE}$ survives, we see that Eq. (\ref{AHE}) is simplified to be 
$j^X_{\rm AHE}(Z)=\frac{e^2}{2\pi h}\frac{U}{M_0}E_Y\sum_f\partial n_f(Z)/\partial Z$.
The total current in the $X$ direction is given by
\begin{align}
\begin{split}
J^X_{\rm AHE}&=\int_0^L dZ j^X_{\rm AHE}(Z)=\frac{e^2}{2\pi h}\frac{UD_2}{M_0}E_Y,
\end{split}
\end{align}
where $D_2=\sum_f\int_{\theta=\theta_0}^{\theta=\theta_0+\delta}dn_f=\sum_f[n_f(\theta_0+\delta)-n_f(\theta_0)]=n_0\{ \sqrt{2}\sin(\varphi_0+\frac{\pi}{4})[\sin(\theta_0+\delta)-\sin\theta_0]+\cos(\theta_0+\delta)-\cos\theta_0 \}$.
The Hall conductivity is estimated as $\sigma_{XY}=\frac{e^2}{2\pi h}\frac{UD_2}{M_0}\sim 1\times 10^{-2}\  e^2/h$, since $Un_0/|M_0|\sim 0.1$ \cite{comment3} and $|D_2|/n_0\sim 1$.
Schematic figure of a possible experimental setup to observe the AHE in our system is shown in Fig. \ref{Fig2}(b).
Two ferromagnets with a relative angle $\delta$ in the magnetization directions are attached to the AFI \cite{Tveten2013}.
In experiments, the $\delta$ dependence of the Hall conductivity will be a direct evidence for the observation of the axion field.
Note that, in contrast to preceding works on the AHE in AFMs \cite{Shindou2001,Chen2014}, the AHE studied here does not occur in uniform ground states.
Namely, spatial variations of the AF order parameter $n_f$ need to be realized by external forces.

{\it Discussions and Summary.---}
Let us discuss briefly the realization of our predictions in realistic correlated systems with SOC.
It has been suggested that the dynamical axion field can be realized by spin excitations in the AFI phases of Bi$_2$Se$_3$ family doped with magnetic impurities \cite{Li2010} and transition metal oxides with corundum structure such as $\alpha$-Fe$_2$O$_3$ \cite{Wang2011}.
In the same manner as above, we can derive similar expressions for the CME and AHE in these systems.
What about the possibility in other systems?
First of all, time-reversal and inversion symmetries of the system must be broken to induce the deviation of $\theta$ from $0$ or $\pi$.
Theoretically, the value of $\theta$ can be calculated numerically in any insulating systems \cite{Essin2009,Essin2010,Coh2011}.
The point is that the emergence of the CME and AHE depends on whether $\theta$ is a function of physical quantities such as AF order parameter, as in our case.
If $\theta$ is a function of a physical quantity, then the fluctuation of the physical quantity realizes the dynamical axion field.
It should be noted that, even if the value of $\theta$ is zero in ground states, the realization of dynamical axion fields is possible.

In summary, we have studied theoretically 3D AFIs with SOC, focusing on a role of the staggered magnetization.
We have revealed that, in the presence of SOC, spin excitations in AFIs can result in nontrivial charge responses.
It is shown that the {\it dynamical} CME, an alternating current generation by magnetic fields, emerges due to the time dependences of the AF order parameter.
It is also shown that the AHE arises due to the spatial variations of the order parameter.
These two phenomena are the consequences of the realization of the dynamical axion field in the AFI phase.
The magnetic-field-induced and electric-field-induced currents in this study are understood as a polarization current in the bulk and a magnetization current in the bulk, respectively, which can flow in insulators.
Observing these phenomena is equivalent to detecting the dynamical axion field in condensed matter.
In other words, we propose a new way to detect the dynamical axion field.

\vspace{1ex}
The authors thank T. Chiba, Y. Araki, O. A. Tretiakov, S. Takahashi, and J. Barker for valuable discussions.
A.S. is supported by a JSPS Research Fellowship.
This work was supported in part by Grant-in-Aid for Scientific Research (No. 26107505 and No. 26400308) from MEXT, Japan.


\begin{widetext}

\vspace{3ex}
\begin{center}
\textbf{{\large Supplemental Material}}
\end{center}
\vspace{3ex}
\textbf{1. Alternative Derivation of Eq. (\ref{S_theta_realtime})}
\vspace{2ex}

In this section, we derive the effective action consisting of the N\'{e}el field $\bm{n}$ and an external electromagnetic potential $A_\mu$.
For this purpose, it is convenient to adopt a perturbative method rather than the Fujikawa's method.
We start with the total action of an antiferromagnetic (AF) insulator described by Eq. (\ref{alpha5-Dirac}) in the presence of an external electromagnetic potential $A_\mu$:
\begin{align}
\begin{split}
S_{\rm eff}[\psi,\bar{\psi},\bm{n},A_\mu]=\int dtd^3 r\sum_f\bar{\psi}_f(\bm{r},t)\left[i\gamma^\mu D_\mu-M_0+i\gamma^5M_{5f}\right]\psi_f(\bm{r},t).\label{Eq-S1}
\end{split}\tag{S1}
\end{align}
By integrating out the fermionic field, we obtain the effective action for $\bm{n}$ and $A_\mu$ as
\begin{align}
\begin{split}
Z[\bm{n},A_\mu]&=\int[\psi,\bar{\psi}]e^{iS_{\rm eff}}=\exp\left\{\sum_f\mathrm{Tr}\ln\left[G_{0f}^{-1}(1+G_{0f}V_f)\right]\right\}=\exp\left[\sum_f\mathrm{Tr}\left(\ln G_{0f}^{-1}\right)+\sum_f\sum_{n=1}^\infty\frac{1}{n}\mathrm{Tr}\left(G_{0f}V_f\right)^n\right]\\
&\equiv e^{iW_{\rm eff}[\bm{n},A_\mu]},\label{Eq-S2}
\end{split}\tag{S2}
\end{align}
where $G_{0f}=(i\gamma^\mu\partial_\mu-M_0)^{-1}$ is the Green's function of the noninteracting part, $V_f=-e\gamma^\mu A_\mu+i\gamma^5M_{5f}$ is the perturbation term, and we have used $i\gamma^\mu D_\mu-M_0+i\gamma^5M_{5f}=G_{0f}^{-1}+V_f$.
Next we use the following identities for the traces of gamma matrices:
\begin{align}
\begin{split}
\mathrm{tr}(\gamma^\mu)=\mathrm{tr}(\gamma^5)=0,\ \ \ \ \mathrm{tr}(\gamma^\mu\gamma^\nu)=4g^{\mu\nu},\ \ \ \ \mathrm{tr}(\gamma^\mu\gamma^\nu\gamma^5)=0,\ \ \ \ \mathrm{tr}(\gamma^\mu\gamma^\nu\gamma^\rho\gamma^\sigma\gamma^5)=-4i\epsilon^{\mu\nu\rho\sigma}.
\end{split}\tag{S3}
\end{align}
The relevant terms up to the leading order read
\begin{align}
\begin{split}
iW_{\rm eff}[\bm{n},A_\mu]=\frac{1}{2}\sum_f\mathrm{Tr}\left(G_{0f}i\gamma^5M_{5f}\right)^2+\sum_f\mathrm{Tr}\left[\left(-G_{0f}e\gamma^\mu A_\mu\right)^2\left(G_{0f}i\gamma^5M_{5f}\right)\right].\label{Eq-S4}
\end{split}\tag{S4}
\end{align}
The first and second terms correspond to a bubble-type diagram and a triangle-type digram, respectively.

For concreteness, we consider the case of $M_{5,1}=Un_1$, $M_{5,2}=Un_2$, and $M_{5,3}=Un_3$ ($U$ is the on-site interaction strength), which is applied to the Fu-Kane-Mele-Hubbard model.
Here the N\'{e}el field is given by $\bm{n}=n_1\bm{e}_x+n_2\bm{e}_y+n_3\bm{e}_z$.
The first term in Eq. (\ref{Eq-S4}) is given explicitly by
\begin{align}
\begin{split}
\mathrm{Tr}\left(G_{0f}i\gamma^5M_{5f}\right)^2&=U^2\int\frac{d^4q}{(2\pi)^4}\int\frac{d^4k}{(2\pi)^4}\frac{\mathrm{tr}\left[ i(\gamma^\mu k_\mu+M_0)i\gamma^5n_f(q)i[\gamma^\nu(k+q)_\nu+M_0]i\gamma^5n_f(-q) \right]}{(k^2-M_0^2)[(k+q)^2-M_0^2]}\\
&=4U^2\int\frac{d^4q}{(2\pi)^4}\int\frac{d^4k}{(2\pi)^4}\frac{[k_\mu(k+q)^\mu+M_0^2]n_f(q)n_f(-q)}{(k^2-M_0^2)[(k+q)^2-M_0^2]}\\
&\equiv \int\frac{d^4q}{(2\pi)^4}I(q)n_f(q)n_f(-q),\label{Eq-S5}
\end{split}\tag{S5}
\end{align}
where $k^2=g^{\mu\nu}k_\mu k_\nu=k_\mu k^\mu=k_0^2-\bm{k}^2$.
We have used $G_{0f}(k)=i(\gamma^\mu k_\mu+M_0)/(k^2-M_0^2)$, $\{\gamma^\mu,\gamma^5\}=0$, and $\{\gamma^\mu,\gamma^\nu\}=2g^{\mu\nu}$.
Furthermore we introduce the Feynman parameter to combine the denominator as
\begin{align}
\begin{split}
\frac{1}{(k^2-M_0^2)[(k+q)^2-M_0^2]}=\int_0^1 d\alpha\frac{1}{[l^2+\alpha(1-\alpha)q^2-M_0^2]^2},
\end{split}\tag{S6}
\end{align}
where $l=k+\alpha q$.
On the other hand, in terms of $l$, we obtain $k_\mu(k+q)^\mu=l^2-\alpha(1-\alpha)q^2+(1-2\alpha)l_\mu q^\mu$.
Then the integral $I(q)$ can be represented as
\begin{align}
\begin{split}
I(q)/i=4U^2\int\frac{d^4l_{\rm E}}{(2\pi)^4}\int_0^1 d\alpha\frac{l_{\rm E}^2+\alpha(1-\alpha)q^2-M_0^2}{[l_{\rm E}^2+M_0^2-\alpha(1-\alpha)q^2]^2}=A+Bq^2+\mathcal{O}(q^4),\label{Eq-S7}
\end{split}\tag{S7}
\end{align}
where we have Wick-rotated as $l^0_{\rm E}=il^0$.
The term that contains $l_\mu q^\mu$ vanishes, because it is an odd function of $l_\mu$.
Finally, by substituting Eq. (\ref{Eq-S7}) into Eq. (\ref{Eq-S5}), we arrive at the action of the form
\begin{align}
\begin{split}
\sum_f\mathrm{Tr}\left(G_{0f}i\gamma^5M_{5f}\right)^2=\frac{i}{g}\int dtd^3r \left[(\partial_\mu \bm{n})\cdot(\partial^\mu \bm{n})+m^2\bm{n}^2\right],\label{Eq-S8}
\end{split}\tag{S8}
\end{align}
where $1/g=B$ and $m^2=A/B$.
This action is nothing but the action of the N\'{e}el field (i.e., the nonlinear sigma model) \cite{Haldane1983}.
In the present low-energy effective model [Eq. (\ref{Eq-S1})], the information on the anisotropy of the N\'{e}el field is not included.
On the other hand, many (actual) AF insulators have the easy-axis anisotropy.
Hence the term $m^2\bm{n}^2$ will be replaced by a term like $m^2(\bm{n}\cdot\bm{e}_A)^2$ with $\bm{e}_A$ denoting the easy axis.

The second term in Eq. (\ref{Eq-S4}) is the so-called triangle anomaly, which gives the theta term.
The final result is \cite{Nagaosa1996,Hosur2010}
\begin{align}
\begin{split}
\sum_f\mathrm{Tr}\left[\left(-G_{0f}e\gamma^\mu A_\mu\right)^2\left(G_{0f}i\gamma^5M_{5f}\right)\right]&=i\int dtd^3r\frac{e^2}{4\pi h}\left[\frac{\pi}{2}[1+\mathrm{sgn}(M_0)]-\sum_f\frac{Un_f(\bm{r},t)}{M_0}\right]\epsilon^{\mu\nu\rho\lambda}\partial_\mu A_\nu \partial_\rho A_\lambda\\
&=i\int dtd^3 r \frac{e^2}{2\pi h}\theta(\bm{r},t) \bm{E}\cdot\bm{B},
\end{split}\tag{S9}
\end{align}
where $\theta(\bm{r},t)=\frac{\pi}{2}[1+\mathrm{sgn}(M_0)]-\sum_fUn_f(\bm{r},t)/M_0$.
Note that this value of $\theta$ is consistent with that obtained by the Fujiawa's method $\theta_{\rm F}=\frac{\pi}{2}[1+\mathrm{sgn}(M_0)]-\sum_f\tan^{-1}(Un_f/M_0)$, since $\tan^{-1}(x)=x-x^3/3+\cdots$.

\vspace{6ex}
\noindent\textbf{2. From the Fu-Kane-Mele-Hubbard Model to the Heisenberg Model}
\vspace{2ex}

In this section, we consider the validity of the dynamics of the sublattice magnetizations described by Eq. (\ref{dynamics}) in the Fu-Kane-Mele-Hubbard model.
A mean-field study has shown that the AF insulator phase of the Fu-Kane-Mele-Hubbard model develops when $U/t\sim 4$ (with $\lambda/t\sim 0.3$) \cite{Sekine-Proc}.
Hence we may consider a strong coupling picture.
When $U\gg t$ and $U\gg \lambda$, we can derive the Heisenberg model by treating the nearest-neighbor (NN) hopping term $H_t=\sum_{\langle i,j\rangle,\sigma}t_{ij}c^\dag_{i\sigma}c_{j\sigma}$ and the next-nearest-neighbor (NNN) spin-orbit coupling term $H_\lambda=4i\lambda/a^2\sum_{\langle\langle i,j\rangle\rangle}c^\dag_{i}\bm{\sigma}\cdot(\bm{d}^1_{ij}\times\bm{d}^2_{ij})c_{j}$ as perturbation terms.
After a standard second-order perturbation calculation, we obtain the NN exchange interaction
\begin{align} 
\begin{split}
H^{\rm NN}_{\rm exch.}=\sum_{\langle i,j\rangle}\sum_{\sigma,\sigma'}\frac{\left[t_{ij}c^\dag_{i\sigma'}c_{j\sigma'}\right]\left[t_{ji}c^\dag_{j\sigma}c_{i\sigma}\right]}{0-U}=\sum_{\langle i,j\rangle}\frac{4t_{ij}^2}{U}\bm{S}_i\cdot\bm{S}_j+\mathrm{const.}
\end{split}\tag{S10}
\end{align}
Similarly, we can obtain the NNN exchange interaction term.
For example, in the case of $\bm{d}^1_{ij}=\frac{a}{4}(1,1,1)$ and $\bm{d}^2_{ij}=\frac{a}{4}(1,1,-1)$, we have $c^\dag_{i}\bm{\sigma}\cdot(\bm{d}^1_{ij}\times\bm{d}^2_{ij})c_{j}=\frac{a^2}{8}c^\dag_i[-\sigma_x+\sigma_y]c_j$.
Then a second-order perturbation reads
\begin{align} 
\begin{split}
H^{\rm NNN}_{\rm exch.}=\frac{\left\{\frac{1}{2}i\lambda c^\dag_i[-\sigma_x+\sigma_y]c_j\right\}\left\{\frac{1}{2}i\lambda c^\dag_j[-\sigma_x+\sigma_y]c_i\right\}}{0-U}=\frac{2\lambda^2}{U}\left(S_i^x S_j^y+S_i^y S_j^x+S_i^z S_j^z\right)+\mathrm{const.}
\end{split}\tag{S11}
\end{align}
We can do the same procedure as above for the other NNN directions.

Combining these things, the exchange interaction term is written as
\begin{align} 
\begin{split}
H_{\rm exch.}=\sum_{\langle i,j\rangle}J_{ij}\bm{S}_i\cdot\bm{S}_j+\sum_{\langle\langle i,j\rangle\rangle}\sum_{a,b}\Lambda^{ab}_{ij}S^a_i S^b_j,\label{Eq-S12}
\end{split}\tag{S12}
\end{align}
where $J_{ij}=4t_{ij}^2/U$ and $\Lambda^{ab}_{ij}=\mathcal{O}(\lambda^2/U)$ ($a,b=x,y,z$).
The fact that $\Lambda^{ab}_{ij}$ contains positive values indicates that the NNN exchange interaction can favor AF alignments between the same sublattices.
Namely, the NNN exchange interaction competes with the NN exchange interaction which favors ferromagnetic alignments between the same sublattices.
Such a competition is consistent with a numerical result from the weak coupling that shows the increase of the critical strength $U_c$ for the AF ordering in the presence of spin-orbit coupling \cite{Sekine-Proc}.
Furthermore, spin-orbit coupling breaks spin SU(2) symmetry, i.e., the orientations of spins are coupled to the lattice structure.
More generally, as well known, the presence of spin-orbit coupling leads to magnetic anisotropy \cite{Yosida-Book}.
Therefore, the effects of the second term in Eq. (\ref{Eq-S12}) can be understood as to modify the value of the NN coupling $J_{ij}$ and to determine the ground-state direction of spins.

In many antiferromagnets, there exists the easy-axis anisotropy such that $H_{\rm aniso.}=-K\sum_i(\bm{S}_i\cdot\bm{e}_{n_0})^2$ with $\bm{e}_{n_0}$ denoting the ground state direction.
For example, in the case of the AF insulator phase of Bi$_2$Se$_3$ doped with magnetic impurities such as Fe, the easy axis is the $z$ direction (perpendicular to the quintuple layers) \cite{Li2010}.
Then, in the presence of an external magnetic field $\bm{B}$, the spin Hamiltonian of the system could be written effectively as
\begin{align} 
\begin{split}
H_{\rm eff}=\tilde{J}\sum_{\langle i,j\rangle}\bm{S}_i\cdot\bm{S}_j-K\sum_i(\bm{S}_i\cdot\bm{e}_{n_0})^2-g\mu_{\rm B}\sum_i\bm{B}\cdot\bm{S}_i,\label{Eq-S13}
\end{split}\tag{S13}
\end{align}
where $\tilde{J}$ includes the effect of spin-orbit coupling and we have assumed isotropic exchange interaction for simplicity.
It should be noted that the continuum action of this Hamiltonian is given by Eq. (\ref{Eq-S8}) (this is the famous Haldane's mapping) \cite{Haldane1983}.

The spin dynamics is determined from the equation of motion:
\begin{align} 
\begin{split}
\dot{\bm{S}}_i=i\left[H_{\rm eff},\bm{S}_i\right]=\bm{S}_i\times\bm{f}_i,
\end{split}\tag{S14}
\end{align}
where $\bm{f}_i$ is the effective magnetic field given by
\begin{align} 
\begin{split}
\bm{f}_i&=-\partial H_{\rm eff}/\partial \bm{S}_i=-\tilde{J}\sum_{\langle i,j\rangle}\bm{S}_j+2K(\bm{S}_i\cdot\bm{e}_{n_0})\bm{e}_{n_0}+g\mu_{\rm B}\bm{B}.
\end{split}\tag{S15}
\end{align}
In order to consider the AF resonance state, where all the spins are precessing around the easy axis with the same frequency, we can replace the spins $\bm{S}_i$ by the mean-field values $\bm{m}_A$ and $\bm{m}_B$ with $A,B$ denoting two sublattices of a diamond lattice.
Finally, we arrive at Eq. (\ref{dynamics}):
\begin{align} 
\begin{split}
\dot{\bm{m}}_A&=\bm{m}_A\times\left\{-\omega_J\bm{m}_B+\left[g\mu_B B+\omega_A(\bm{m}_A\cdot\bm{e}_{n_0})\right]\bm{e}_{n_0}\right\},\\
\dot{\bm{m}}_B&=\bm{m}_B\times\left\{-\omega_J\bm{m}_A+\left[g\mu_B B+\omega_A(\bm{m}_B\cdot\bm{e}_{n_0})\right]\bm{e}_{n_0}\right\},
\end{split}\tag{S16}
\end{align}
where $\omega_J=\tilde{J}z_{\rm NN}$ ($z_{\rm NN}$ is the number of NN bonds), $\omega_A=2K$, and a static magnetic field is applied along $\bm{e}_{n_0}$ as $\bm{B}=B\bm{e}_{n_0}$.
Note that the easy-axis anisotropy term in Eq. (\ref{Eq-S13}) is {\it not} essential to cause the dynamics.
What is essential is an external magnetic field, as is understood from the resonance frequency $\omega=\omega_\pm=g\mu_B B\pm\sqrt{(2\omega_J+\omega_A)\omega_A}$.

\end{widetext}

\begin{thebibliography}{99}
\bibitem{comment-B}
The magnetic-field-induced and electric-field-induced terms are understood as a polarization current in the bulk and a magnetization current in the bulk, respectively, which can flow in insulators.
Namely, they are obtained as $\partial\bm{P}/\partial t=e^2/(2\pi h)\dot{\theta} \bm{B}$ and $\nabla\times\bm{M}=e^2/(2\pi h)\nabla\theta\times \bm{E}$, where we have considered a case of static electromagnetic fields $\bm{B}$ and $\bm{E}$.
\bibitem{comment3}
Since we assume that the system lies in the vicinity of the phase boundary, the condition $Un_f/M_0\ll 1$ is needed.
\end{thebibliography}
\end{document}